# Science drivers and requirements for an *Advanced Technology Large Aperture Space Telescope* (ATLAST): Implications for technology development and synergies with other future facilities

Marc Postman[*a], Tom Brown[a], Kenneth Sembach[a], Mauro Giavalisco[b], Wesley Traub[c], Karl Stapelfeldt[c], Daniela Calzetti[b], William Oegerle[d], R. Michael Rich[e], H. Phillip Stahl[f], Jason Tumlinson[a], Matt Mountain[a], Rémi Soummer[a], Tupper Hyde[d]

[a]Space Telescope Science Institute, 3700 San Martin Drive, Baltimore, MD USA 21218;
[b]University of Massachusetts, Dept. of Astronomy, Amherst, MA USA 01003;
[c]Jet Propulsion Laboratory/California Institute of Technology, Pasadena, CA USA 91109;
[d]Goddard Space Flight Center, Greenbelt, MD USA 20771;
[e]University of California, Division of Astronomy, Los Angeles, CA USA 90095;
[f]Marshall Space Flight Center, MS SD70 SOMTC, Huntsville, AL USA 35812-0262

**ABSTRACT**

The Advanced Technology Large-Aperture Space Telescope (ATLAST) is a concept for an 8-meter to 16-meter UVOIR space observatory for launch in the 2025-2030 era. ATLAST will allow astronomers to answer fundamental questions at the forefront of modern astronphysics, including "Is there life elsewhere in the Galaxy?" We present a range of science drivers that define the main performance requirements for ATLAST (8 to 16 milliarcsec angular resolution, diffraction limited imaging at 0.5 μm wavelength, minimum collecting area of 45 square meters, high sensitivity to light wavelengths from 0.1 μm to 2.4 μm, high stability in wavefront sensing and control). We will also discuss the synergy between ATLAST and other anticipated future facilities (e.g., TMT, EELT, ALMA) and the priorities for technology development that will enable the construction for a cost that is comparable to current generation observatory-class space missions.

**1. Introduction**

The most compelling astrophysical questions to be addressed in the 2020 era will, like those today, be pursued using data obtained from both space-based and ground-based telescopes. The impressive capabilities anticipated for ground-based observatories in the upcoming decade (e.g., 20 to 40-meter class optical telescopes, ALMA, and potentially the square kilometer array (SKA) near the end of the decade) will redefine the existing synergy between ground and space telescopes. Advances, over the next decade, in Multi-Conjugate Adaptive Optics (MCAO) and Ground-Layer Adaptive Optics (GLAO) for large aperture ground-based telescopes[1,2,3,4] may enable intermediate to high Strehl ratio (~40-80%) performance over fields of view of perhaps up to 2 arcminutes across for wavelengths longwards of ~1 μm. Advances in Extreme AO[5,6,7] may enable high Strehl ratio performance down to wavelengths as short as 0.55 μm but only over very small fields of view of 1 or 2 arcseconds. Space-based telescopes, however, are the optimal facilities for observations that require any combination of very high-angular resolution and precise wavefront control over fields of view larger than ~2 arcminutes or at all wavelengths shorter than 1 μm, very high sensitivity (nanoJansky levels), very stable PSF performance across the field of view, high photometric precision (< 0.0001 mag) and accuracy in crowded fields, and very high stability of all these performance parameters over tens to hundreds of hours of exposure time.

---

[*] Email: postman@stsci.edu; phone (410) 338-4340

Several independent scientific drivers that we will highlight here require primary aperture diameters of at least 8 meters and, for some applications, as large as 16 meters. The next large UVOIR space telescope must be diffraction-limited at least down to 0.5 µm and must have good sensitivity down to 0.11 µm. A space telescope with such capability will be revolutionary because it enables fundamental breakthroughs in astrophysics – both on its own and in combination with other telescopes with different capabilities. We have studied several designs for this next generation space telescope, a suite of concepts we collectively refer to as the Advanced Technology Large-Aperture Space Telescope (ATLAST) [8,9,10,11]. ATLAST has the performance required to detect the potentially rare occurrence of biosignatures in the spectra of terrestrial exoplanets, to reveal the underlying physics that drives star formation, and to trace the complex interactions between dark matter, galaxies, and the intergalactic medium.

The challenge the astronomy community faces, given the compelling science enabled by ATLAST, is how to construct it for a cost that is comparable to current flagship-class missions. Modern ground-based telescopes (post-1992) follow an aperture size – cost relationship that is significantly shallower than older designs, owing to the advance of technology. Similar trends are seen for space-based telescopes[12,13]. The construction costs for the James Webb Space Telescope (JWST) are similar to those of the Hubble Space Telescope (HST) and yet JWST has an aperture that is 2.7 times larger and a mass that is a factor of 1.7 smaller. Key technologies that can continue to flatten the slope of the space telescope aperture-cost relationship include lightweight mirror materials and fabrication methods[14,15,16], closed-loop wavefront control of active optics, disturbance isolation systems, modular design, high efficiency UV detectors, and ultra-low noise optical/IR detectors. If significant investments are made in these technologies, it is plausible that 8-meter and larger UVOIR space telescopes would be affordable by NASA in the 2020 era. An important element to achieving maturity of the technology needed for a large UVOIR (non-cryogenic) space telescope is partnership with other communities that have similar technology drivers (e.g., national reconnaisance, defense, and remote sensing of Earth).

## 2. Astrophysical Science Drivers

Conceptual breakthroughs in understanding astrophysical phenomena are made when our ability to probe structures on the relevant angular scales is enabled by our astronomical observatories. HST tremendously advanced our understanding of galaxy evolution because its angular resolution (~65 mas) at its diffraction limit (0.63 µm) is, for *all* redshifts, less than or equal to one half of the angular scale of the kiloparsec-size features within galactic systems that reveal key morphological information. ATLAST will be poised to make equally large scientific breakthroughs if it can achieve spatial resolutions that are more than 4 times better than HST and JWST. Achieving ultra-high dynamic range (high-contrast) imaging and/or long-term imaging with high spatial and temporal PSF stability at angular resolutions of ~8 to 15 mas in the 0.11 µm to 2.4 µm wavelength range are required for some extremely compelling science that will not be readily achieved by any other facility. For example, various combinations of high spatial resolution, high sensitivity, and high contrast are needed if we wish to definitively detect the potentially rare occurrence of biosignatures in the atmospheres of terrestrial–mass exoplanets, or to study stellar populations across the full range of galaxy environments, or to trace the kinematics of gas and dark matter on galactic scales to directly map the growth of structure over time. We discuss these cases in brief detail below. These are just a few of the many exciting investigations requiring a large UVOIR space telescope (and which will enable an equal or greater number of as yet unimagined discoveries).

### 2.1 Terrestrial exoplanet characterization: "Are We Alone?"

We are at the brink of answering two paradigm-changing questions: Do Earth-sized planets exist in the Habitable Zones of their host stars? Do any of them harbor life? The tools for answering the first question already exist (e.g., Kepler, CoRoT); those that can address the second can be developed within the next 10-20 years[17]. ATLAST is our best option for an extrasolar life-finding facility and is consistent with the long-range strategy for space-based initiatives recommended by the AAAC Exoplanet Task Force[18]. Four significant drivers dictate the need for a large space-based telescope if one wishes to conduct a successful search for biosignatures on exoplanets. First, and foremost, Earth-mass planets are faint – an Earth twin at 10 pc, seen at maximum elongation around a G-dwarf solar star, will have V ~ 29.8 AB mag. Detecting a biosignature, such as the presence of molecular oxygen in the exoplanet's atmosphere, will require the ability to obtain direct low-resolution spectroscopy of such extremely faint sources. Second, the average projected angular radius of the Habitable Zone (HZ) around nearby F,G,K stars is less than 100 milli-arcseconds (mas). One thus needs an imaging system capable of angular resolutions of ~10 to 25 mas to adequately sample the HZ and isolate the exoplanet point source in the presence of an exo-zodiacal background. Third, direct detection of an Earth-sized planet in the HZ requires high contrast imaging, typically requiring starlight suppression factors of $10^{-9}$ to $10^{-10}$. Several techniques [19] are, in principle, capable of delivering such high contrast levels but all require levels of wavefront stability not possible with ground-based telescopes. Guyon[20] and Mountain et al.[21] show that there is a limit to the achievable dynamic range in high contrast imaging from the ground. This is because any AO system is fundamentally limited by the capacity to analyze the wavefront due to the finite number of photons available for wavefront sensing. Even with the 30 to 40 m class telescopes expected to come on line later this decade, this limit corresponds to a contrast ratio of ~$10^{-8}$ (see Figure 4 in Mountain et al.[21]). A space-based platform is thus required to achieve the wavefront stability that is needed for such high contrast imaging of terrestrial mass planets in the HZ. Lastly, biosignature-bearing planets may well be rare, requiring one to search tens or even several hundred stars to find even a handful with compelling signs of life. Therefore if we have to survey a number of stars, and hence limit the exposure time on each candidate, the number of stars for which one can obtain an exoplanet's spectrum at a given SNR scales approximately as $D^3$, where D is the telescope aperture diameter. This is demonstrated in Figure 1 where we have averaged over different simulations done using various starlight suppression options (internal coronagraphs of various kinds as well as an external occulter). To estimate the number of potentially habitable worlds detected, one must multiply the numbers in Figure 1 by the fraction of the stars that have an exoplanet with detectable biosignatures in their HZ ($\eta_{EARTH}$). The value of $\eta_{EARTH}$ is currently not constrained but it is not likely to be close to unity. One must conclude that to maximize the chance for a successful search for life in the solar neighborhood requires a space telescope with an aperture size of at least 8 meters.

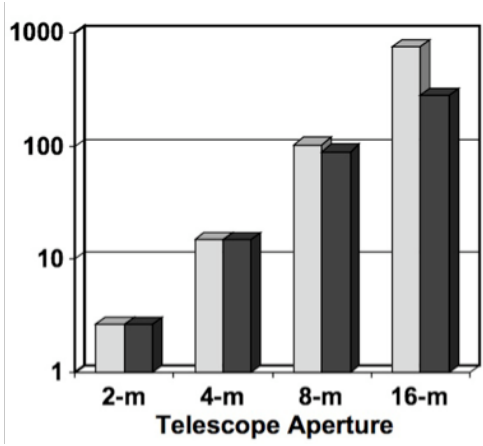

**Figure 1**: The number of spectral type F,G,K stars as a function of telescope aperture where an R=70 SNR=10 spectrum could be obtained of an Earth-twin in the HZ in less than 500 ksec. Light grey shows total number of stars that could be observed at least once. Dark grey shows number of stars that could be visited 3 times in 5 years without exceeding 20% of available telescope time. It is assumed every star has an Earth twin.

Figure 2 shows two simulated ATLAST spectra for an Earth-twin at 10 pc, one at R=100 and one at R=500, taken with sufficient exposure to reach SNR=10 at 0.75 μm in the continuum. A 3-zodi background was used (local plus exosolar). For these calculations we use a fully validated model of the Earth's spectrum[22,23] in combination with the observed visible reflection spectrum of the present Earth. We assume that the exoplanet is at maximum elongation. The R=100 exposure times are 46 ksec and 8 ksec, respectively, for an 8-m and 16-m space telescope. The corresponding exposure times for the R=500 spectrum are 500 ksec and 56 ksec, respectively, for the 8-m and 16-m telescopes. The reflected flux from an Earth-like rocky planet increases as $M^{2/3}$, where M is the exoplanet mass. Hence, the exposure times for a 5 Earth-mass exoplanet would be ~3 times shorter. At both resolutions, the $O_2$ features at 0.68 μm and 0.76 μm are detected, as are the $H_2O$ features at 0.72, 0.82, 0.94, 1.13, 1.41, and 1.88 μm. Rayleigh scattering is detected as an increase in reflectivity bluewards of 0.55 μm. The higher spectral resolutions enabled by large-aperture space telescopes enable the detection of molecular oxygen in exoplanets with lower abundances than those on Earth and provide constraints on the kinematics and thermal structure of the atmosphere that are not accessible at lower resolution.

For a 16-m class space telescope, time-resolved spectroscopy over intervals of a few hours may reveal surface composition variations, if the planet is not cloud dominated, as the exoplanet rotates. However, even broadband photometry can be used to detect short-term variations in albedo that can determine the rotation period and constrain the amount of cloud cover. ATLAST will allow us to glean substantial information about an exo-Earth from temporal variations in its features. Such variations inform us about the nature of the dominant surface features, changes in climate, changes in cloud cover, and potentially, seasonal variations in surface vegetation. Ford et al.[24] generated model light curves for the Earth over 6 consecutive days using data from real satellite observations. Photometric variations of 20 – 30% on timescales of 6 hours were typical in the B,V,R,I passbands. To track such variations with a SNR of 20 on an Earth-like planet (with a similar rotation period as Earth) at a distance of 20 pc would require a space telescope with an aperture of at least 8 meters. A 4-meter space telescope would be able to perform such observations only for planetary systems within 10 pc. As the number of terrestrial planetary systems scales as the cube of the distance, the ability to reach to 20 pc provides nearly an order of magnitude more targets.

The instrumentation required to perform the above observations on an Earth-twin at up to 20 pc distance requires a starlight suppression system that allows detection of exoplanets that are ~25 magnitudes fainter than their host star at an inner working angle of ~40 mas. This is the baseline mission starlight suppression performance requirement. There are several options for the suppression system – an internal coronagraph or external occulter. Achieving the above level of suppression with a segmented telescope will require development of a nulling coronagraph and/or an external occulter (starshade)[1]. While the design of the external occulter is independent of the telescope's optical design, the viable options for an internal coronagraph are dependent upon the telescope's optical design: a Lyot or masked-based coronagraph can be used with telescope that employs an off-axis secondary mirror (SM) and a monolithic primary mirror (PM) and, possibly, with one that has an on-axis SM and monolithic PM if the SM is supported by a single linear structure. However, for an on-axis SM with standard spider supports or for a segmented PM, the only internal coronagraph concept that would, in principle, do the job is a visible nulling coronagraph (VNC). As part of the early technology development plan for ATLAST, these options would be investigated and a downselect made prior to entry into phase A.

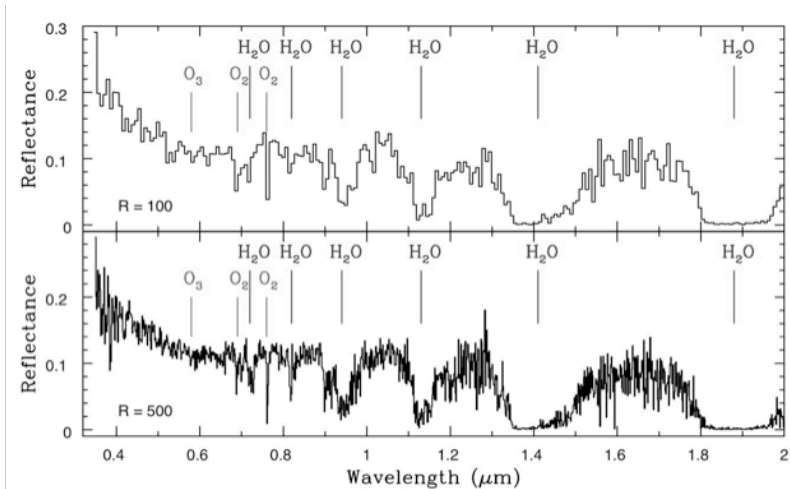

**Figure 2: Simulated ATLAST spectra of an Earth-twin at 10 pc, shown at R=100 (top) and R=500 (bottom). The SNR=10 at 0.75 micron in both cases. Key molecular oxygen and water features are shown. Increased reflectcance at the blue end is due to Rayleigh scattering.**

An 8-m ATLAST (with an internal coronagraph) will be able to observe ~100 different star systems 3 times each in a 5-year interval and not exceed 20% of the total observing time available to the community. A 16-m version (with an internal coronagraph) could visit up to ~250 stars three times each in a five-year period. ATLAST used in conjunction with a single external occulter can observe ~85 stars 3 times each in a 5-year period, limited by the transit times of the occulter.

The impact on the ATLAST science objectives if a starlight suppression system is not able to permit detection of sources at ~$10^{-10}$ contrast ratio at the indicated inner working angle (IWA) would be the inability to **directly** detect and characterize Earth-mass planets around solar-type stars (indirect characterization via transit spectroscopy would be unaffected). Characterization of super-Earths (up to 10x Earth's mass) or Earth-like planets around M-dwarfs would still be possible, however, if contrasts of $10^{-9}$ are reachable. We adopt $10^{-9}$ as the minimum mission criterion for starlight suppression performance. The ATLAST science flowdown for terrestrial exoplanet characterization is shown in Table 1.

---

[1] A thorough discussion of starlight suppression for ATLAST is available in the appendices of our public NASA study report at http://www.stsci.edu/institute/atlast.

Table 1. ATLAST Science Flowdown Requirements for Biosignature Detection on Exoplanets

| Science Question | Science Requirements | Measurements Needed | Design and Implementation |
|---|---|---|---|
| **Is there life elsewhere in the Galaxy?** | Detect at least 10 Earth-like Planets in HZ with 95% confidence if $\eta_{EARTH} = 0.15$ | High contrast ($\Delta Mag > 25$ mag) SNR=10 broadband (R=5) imaging with IWA ~ 40 mas for ~100 target stars. | At least 8 m PM aperture to achieve sample size (~100 stars) and the desired SNR spectra in <500 ksec for most distant (~25 pc) exoplanets in the sample. Stable $10^{-10}$ starlight suppression on detector at IWA. |
| | Detect the presence of habitability and bio-signatures in the spectra of Earth-like HZ planets | High contrast ($\Delta Mag > 25$ mag) SNR=10 low-resolution (R=70-100) spectroscopy with an IWA ~ 40 mas. Exposure times <500 ksec. | 0.1 nm WFE at 0.6 μm over 2 hr; ~1.3 to 1.6 mas pointing stability. IFU with sensitivity from 0.3 to 2.4 μm, with broadband imaging and two spectroscopic modes (R=70, R=500). |

## 2.2 Stellar population characterization

ATLAST will, for the first time, enable the reconstruction of complete star formation histories (spanning 10 Gyr) for hundreds of galaxies beyond the Local Group, opening the full range of star formation environments to exploration. A comprehensive and predictive theory of galaxy formation and evolution requires that we accurately determine how and when galaxies assemble their stellar populations, and how this assembly varies with environment. By definition the dwarf galaxies we see today are not the same as the dwarf galaxies and proto-galaxies that were disrupted during assembly. Our only insight into those disrupted building blocks comes from sifting through the resolved field populations of the surviving giant galaxies to reconstruct the star formation history, chemical evolution, and kinematics of their various structures[25]. *Resolved stellar populations are cosmic clocks*. Their most direct and accurate age diagnostic comes from observations that can resolve the individual, older stars that comprise the main sequence turnoff. But the main sequence turnoff rapidly becomes too faint to detect with any existing telescope for any galaxy beyond the Local Group. This greatly limits our ability to infer much about the details of galactic assembly because the galaxies in the Local Group are not representative of the galaxy population at large. ATLAST will allow us to reach well beyond the Local Group as shown in Figure 3. HST and JWST cannot reach any large galaxies besides our Milky Way and M31 because they lack the required angular resolution. An 8-meter space telescope can reach 10 Gyr old stars in 140 galaxies including 12 giant spirals and the nearest giant elliptical. A 16-meter space telescope extends our reach to the Coma Sculptor Cloud, netting a total of 370 galaxies including 45 giant spirals and 6 ellipticals. Deriving ages and other galactic properties from color-magnitude data requires photometry for thousands of stars spanning 4 orders of magnitude in luminosity. Such observations require a wide-field imager on ATLAST with half-Nyquist sampling over a field of view of at least 4 arcminutes. The ATLAST science flowdown for star formation history reconstruction is shown in Table 2.

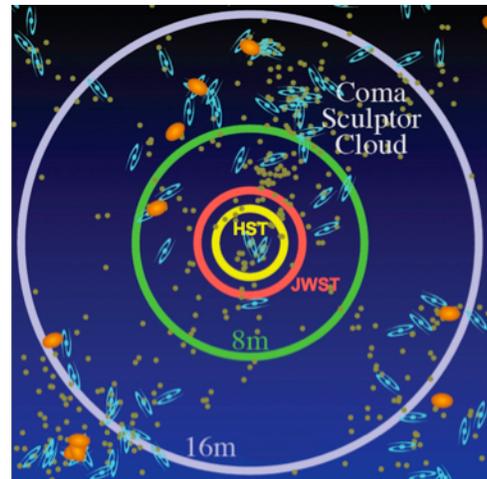

**Figure 3: Nearby galaxy distribution, color-coded by morphological type (blue = spiral, orange = elliptical, olive: dwarf galaxy). Large circles show how far away a solar luminosity star can be detected at SNR=5 in V and I band in a 100 hour exposure.**

ATLAST will work in concert with 30m-class ground-based telescopes (e.g., TMT), expanding our reach to other well-populated galaxy groups, with ATLAST obtaining photometry of V~35 magnitude G dwarf stars and TMT obtaining kinematics of much brighter giants out to the Coma Sculptor Cloud. The dwarf stars in the Coma Sculptor Cloud are Postman et al.: ATLAST Science Driverseffectively inaccessible to TMT, requiring gigaseconds of integration even for an isolated star. Ground-based telescopes and ATLAST will also complement each other in establishing the universality of the local initial stellar mass function (IMF) – a fundamental observable that must be predicted by any viable comprehensive theory of star formation. Large ground-based telescopes with NIR AO-enabled imaging will establish the universality of the low-mass (<2 $M_{SUN}$) end of the IMF and ATLAST will definitively determine the same for the high-mass end of the IMF, which can only be directly measured in UV / optical.

Table 2. ATLAST Science Flowdown Requirements for Star Formation History Reconstruction

| Science Question | Science Requirements | Measurements Needed | Design and Implementation |
|---|---|---|---|
| **What are the star formation histories of galaxies?** | Determine the ages and metallicities of stellar populations over a broad range of galactic environments. Accuracy requirements: age bins ~1 Gyr over lifetime of galaxy; metallicity bins of ~0.2 dex over full range of abundances. | Color-magnitude diagrams using broadband imaging (SNR=5) for individual solar analog stars (absolute V mag ~ 5) in spiral, lenticular, and elliptical galaxies. Imaging must be done in at least two passbands, with bluer band below 0.6 μm wavelength. | At least 8 m PM aperture to observe nearest giant elliptical galaxy and get SNR in 2 passbands in a total exposure <400 ksec.<br><br>WFE: Diffraction limited at 0.5 μm; ~1.3 to 1.6 mas pointing stability; Symmetric PSF highly desirable<br><br>**VIS/NIR wide-field (4 - 8 arcmin FOV) imager** at TMA focus for simultaneous photometry of >10,000 stars |

**2.3 Galaxy Halo and Gas Physics Revealed in Unprecedented Detail**

The ultraviolet (UV) region of the electromagnetic spectrum is highly sensitive to many fundamental astrophysical processes and, hence, measurements at rest-UV wavelengths provide robust, and often unique, diagnostics of the roles of these processes in a establishing a variety of astronomical environments and in controlling the evolution of a variety of objects. Our knowledge of star formation and evolution, of the growth of structure in the universe, of the physics of jet phenomena on many scales, of the nature of AGN, of aurora on and atmospheric composition of the gas giant planets, and of the physics of protoplanetary disks has either been gleaned or greatly expanded through UV observations.

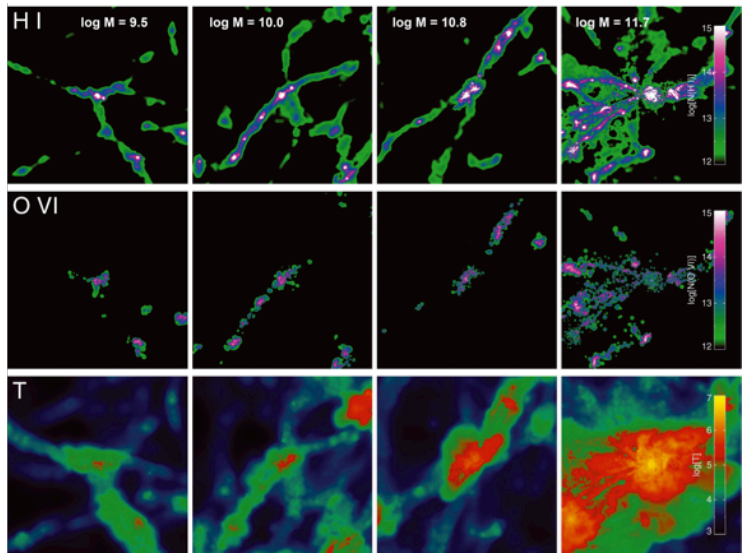

**Figure 4:** Simulations of the gas density (as a function of ionization state) and temperature distribution for 4 different size galaxies (Oppenheimer et al. 2009 [26]). ATLAST will map these distributions.

There is great scientific power in combining high spatial resolution with sensitive UV spectrographic capabilities. One very important application is in the area of galaxy formation. We know that galaxies form and evolve but we know little about how this happens. The physical processes involve complex interactions between the baryonic matter in the galaxies, the energy exchanged during the birth and death of stars, the gas outside the galaxies in the intergalactic medium (IGM), other neighboring galaxies, and the dark matter that dominates and shapes the underlying gravitational potential. Revealing the physics behind galaxy formation and evolution requires making a broad array of observations from the current epoch to the epoch of the first stars. By enabling deep and extensive probes of the IGM in the UV and optical regime, ATLAST will provide many of the key pieces needed to solve this puzzle, particularly in the redshift range $z < 3$ when the cosmic star formation rate peaks and then fades, and galaxies develop their current morphologies.

Understanding how gas in the IGM gets into galaxies and how galaxies respond to inflow lies at the heart of understanding galactic evolution. The mode of accretion depends on the depth of the potential well (galaxy type) and the location at which the intergalactic gas is shocked as it encounters that potential[27,28]. Depending on the mass of the galaxy halo, the infalling gas may be shocked and heated or accrete in "cold mode" along narrow filaments. Gas can also be removed from galaxies via tidal and ram pressure stripping, or during the accretion of gas-rich dwarfs onto giant galaxies. Metal-enriched gas introduced into the IGM by these processes will be dynamically cool. All of these accretion and gas removal theories have observational consequences that can be tested if the properties of gas (e.g., temperature, density, velocity dispersion, metallicity) in and around galaxies can be characterized through absorption and emission line spectroscopy. Figure 4 shows the variation in these properties that models predict as a function of galaxy halo mass. Access to UV wavelengths is required to observe the (slightly redshifted) diagnostic lines (e.g., OVI, SiIII, Ly$\alpha$, NV, SiIV, CIV) needed to characterize the warm IGM at low redshift or OIII, OV, NIV, NeVIII for characterization of the intermediate redshift ($0.3 < z < 2$) IGM. The observational challenge is to acquire datasets of sufficient spatial sampling and with enough diagnostic power (i.e., spectral resolution) to identify and characterize gas in galactic halos. The key requirement is to observe a sufficient number of background sources around each galaxy as well as a large range in galaxy parameter space (e.g., mass, morphology, star formation rate, redshift). Covering a broad range in all parameters is currently beyond the capability of HST. ATLAST will have sufficient UV absorption line sensitivity to be able to survey up to ~100 quasars per deg$^2$ (corresponds to a flux limit of ~1 x 10$^{-17}$ erg cm$^{-2}$ sec$^{-1}$ Å$^{-1}$) over an area that subtends ~10 Mpc on a side (0.25 deg$^2$ at $z = 0.4$) and obtain SNR=10-20 high-resolution (R=20,000) spectra of each QSO in the region. At this limit, ~20% of randomly selected fields on the sky would have a sufficient number of background sources to enable detailed mapping of the spatial distribution of IGM structure around specific galaxies. ATLAST's large aperture coupled with efficient (~40% QE) UV detectors and an efficient spectrograph are needed to enable an individual QSO observation to be completed in ~25 ksec (which permits a 0.25 deg$^2$ area to be surveyed in ~1 week).

Table 3. ATLAST Science Flowdown Requirements for IGM Characterization

| Science Question | Science Requirements | Measurements Needed | Design and Implementation |
|---|---|---|---|
| **How do galaxies and the IGM interact? How does this interaction effect galaxy evolution?** | Map, at high spatial sampling, the properties and kinematics of the intergalactic medium, over contiguous regions of the sky, on scales up to ~10 Mpc. | SNR=20 high resolution (R=20,000) UV spectroscopy of quasars down to FUV mag = 24. | High efficiency UV detectors with at least an 8 m PM aperture to survey wide areas in less than 2 weeks of time.<br><br>**Hi-res UV/blue spectrograph** at Cass focus with sensitivity from 0.11–0.4 μm. |

The dramatically increased absorption line sensitivity at UV and optical wavelengths of ATLAST is crucial for reaching the required background source densities. At the required sampling given above, one can select sight lines next to thousands of examples of any common galaxy, group, or cluster. ATLAST could then be used to produce a high-resolution map of the gas and metals surrounding these structures, which could be used to compare directly against simulation predictions[29,30]. With ATLAST one could also use multiple quasars *and* distant galaxies as background continuum sources to dissect the gas distribution in fields known to have galaxies and gas at the same redshift[31]. ATLAST's large aperture will enable contiguous regions of ~10 Mpc on a side (like in Figure 4) to be surveyed in about 2 weeks of exposure time (with an 8-m ATLAST with enhanced UV detectors) or in ~3 days (with a 16-m ATLAST with enhanced UV detectors)[32]. ATLAST could also be used systematically to target individual nearby galactic coronae and groups of galaxies, for which it would be possible to observe the production sites of heavy elements (star-forming regions, SNe, emission nebulae), follow the processes by which the elements are transported within galaxies, and trace the expulsion of metals into the IGM. The ATLAST science flowdown for IGM characterization is shown in Table 3.

Large ground based telescopes will probe IGM kinematics and structure at higher redshifts where the key spectral features get redshifted into optical or near-IR. However, the increasingly ubiquitous Ly-alpha absorption at higher z is challenging. Hence, the combination of a space-based survey covering z < 2 regime and ground-based covering z>2 makes for an ideal combination.

Recent observations with the Cosmic Origins Spectrograph (COS) on Hubble have shown that it is now possible to detect million-degree intergalactic gas in the ultraviolet lines of O VI and H I Ly-alpha[33]. Studies of intergalactic gas at these temperatures have been the province of x-ray observatories, but may now be broadened significantly to include dedicated studies with future UV-optical telescopes in space. The higher resolution (10-100x) and higher sensitivity (100-1000x) achievable with UV-optical observatories such as ATLAST will at last permit the observational study of this important component of the hot IGM in the context of galaxy environments for large samples (100s-1000s) of intergalactic sight lines. Hence, joint observations with ATLAST and an x-ray observatory will provide very powerful constraints on the hot gas distribution.

**2.4 Dark Matter Dynamics**

Dwarf spheroidal galaxies (dSph), the faintest galaxies known, are extraordinary sites to explore the properties of non-baryonic dark matter (DM). There are several reasons for this. First, their mass is dominated by DM – they are observed to have mass-to-light ratios 10 to 100 times higher than the typical L* galaxy, such as M31 or the Milky Way[34,35,36]. Second, they are relatively abundant nearby – to date ~40 dSph galaxies have been found in the Local Group and more will be discovered. Third, and perhaps most striking, is the discovery that all nineteen dSph satellites of the Milky Way, covering more than four orders of magnitude in luminosity, inhabit dark matter halos with the *same* mass (~$10^7$ $M_{SUN}$) within their central 300 pc[37]. The ability of DM to cluster in phase space is limited by intrinsic properties such as mass and kinetic temperature. Cold dark matter particles have negligible velocity dispersion and very large central phase-space density, resulting in cuspy density profiles. Warm dark matter halos, in contrast, have smaller central phase-space densities, so that density profiles saturate to form constant central cores. Owing to their small masses, dSphs have the highest average phase space densities of any galaxy type, and this implies that for a given DM model, phase-space limited cores will occupy a larger fraction of the virial radii. Hence, the mean density profile of dSph galaxies is a fundamental constraint on the nature of dark matter.

Current observations are unable to measure the density profile slopes within dSph galaxies because of a strong degeneracy between the inner slope of the DM density profile and the velocity anisotropy of the stellar orbits. Radial velocities alone cannot break this degeneracy even if the present samples of radial velocities are increased to several thousand stars[37]. Combining proper motions with the radial velocities is the only robust means of breaking the anisotropy-inner slope degeneracy. The required measurements include proper motions for ~100 stars per galaxy with accuracies better than 10 km/sec (< 40 µas/yr at 60 kpc) and ~1000 line-of-sight velocities. In the case of the brightest of these dSph galaxies, such as Fornax and Sculptor, sufficient velocities and proper motions can be obtained using stellar giants. ATLAST, however, can perform the astrometric measurements. To accomplish this, ATLAST will measure transverse stellar velocities to an accuracy of 5 km/sec. At a distance of 50 kpc, this corresponds to an angular displacement of 0.1 mas over 5 years (about 40 times better than what HST can currently measure[38]). This is approximately one two-hundredths of a pixel, or equivalently, one two-hundredth of the FWHM of the point spread function (PSF). For reference, the Advanced Camera for Surveys (ACS) on HST has centroiding errors of about one-hundredth of the pixel and PSF (the pixel size and PSF have nearly identical widths). However, HST suffers from large thermal stresses on orbit, which cause significant changes in the lengths, positions and alignment of the supporting structures. At the Sun-Earth second Lagrange point (SEL2), ATLAST is far more thermally stable, and the sensors and actuators put in place to maintain the structure to the precision necessary for exoplanet science allows ATLAST to achieve one-sigma astrometric errors of 0.005 pixels. In addition to the instrumental capability for astrometry to 0.1 mas, background objects are needed to provide a stable astrometric reference frame. Quasars alone are likely to be too sparse to provide this frame, so they will be supplemented with background galaxies. While an individual galaxy is far less valuable as an astrometric source than a quasar of a similar magnitude, if the imager used for this investigation has a FOV wide enough to contain thousands of galaxies in a single exposure, then their internal structures, which will be resolved down to 15 mas, will provide the required reference frame. These observations require a wide-field imager on ATLAST with a field of view of ~5 arcminutes. This experiment is not signal-to-noise limited, as solar mass stars at 50 kpc will have signal-to-noise ratios exceeding 100 in ~1 ksec exposures. Thus, even in some of the faintest known dwarf spheriodals, with mass to light ratios exceeding 10,000, the transverse velocity measurements for the 200+ stars per dwarf required by this experiment can be obtained. The dark matter kinematics science flowdown is shown in Table 4.

Ground-based 8-m class telescopes could measure the spectra, and SIM could measure the proper motions for the nearest dwarfs. For the less massive dwarfs, where the dark matter dominance is the greatest, main sequence stars will have to be used to obtain the numbers of velocity and proper motion measurements needed. This will require larger (30-m class) ground-based telescopes for the velocities. The 30-m class telescopes may also be able to obtain the necessary proper motions but it will be extremely challenging: it will require precisely stitching many fields together, most of which are unlikely to contain background quasars of sufficient brightness to be useful as astrometric references. However, the necessary astrometric precision would be readily achieved with ATLAST, given its comparatively wide field of view and stability. ATLAST will, thus, provide some of the best constraints on the nature of dark matter.

Table 4. ATLAST Science Flowdown Requirements for Dark Matter Kinematics

| Science Question | Science Requirements | Measurements Needed | Design and Implementation |
|---|---|---|---|
| **What are the kinematic properties of Dark Matter** | Determine the mean mass density profile of high M/L dwarf Spheroidal Galaxies | Proper motions of ~200 stars per galaxy with accuracies ~20 μas/yr at 50 kpc (yielding required transverse velocity accuracy of 5 km/sec). Augment with stellar radial velocities from ground. Need wide-FOV to ensure sufficient number of background astrometric references (e.g., galaxies and QSO). | Aperture diameter: ≥8 m to achieve the angular resolution to enable stellar centroids to be determined to 0.1 mas. Focal plane metrology must be maintained to enable 0.005 pixel centroid accuracy over a 5-year baseline.<br><br>WFE: Diffraction limited at 0.5 μm; ~1.3 to 1.6 mas pointing stability; symmetric PSF is essential.<br><br>**VIS/NIR wide-field (5 - 8 arcmin FOV) imager at TMA focus.** |

## 3. Technology Development

All ATLAST concepts require many of the same key technologies because they have several fundamental design features in common. All ATLAST concepts are designed to operate in a halo orbit at the Sun-Earth L2 point. The optical designs are diffraction limited at 0.5 μm (36 nm RMS WFE) and the optical telescope assembly (OTA) operates near room temperature (280 K – 290 K). All OTAs employ two simultaneously usable foci: a three-mirror anastigmat (TMA) channel for multiple, wide field of view instruments and a Cassegrain channel for high-throughput UV instruments and instruments for imaging and spectroscopy of exoplanets. All designs have an RMS WFE of < 5 nm at < 2 arcseconds radial offset from Cassegrain optical axis. See the ATLAST concept study[8,9] and also Stahl et al., Oegerle et al., and Hopkins et al. (these proceedings) for additional details.

### 3.1 Telescope Technology

The diffraction-limited imaging at 0.5 μm that is needed for much of ATLAST science requires HST-quality mirror surface errors (5-10 nm rms) to meet the overall system wavefront error of 36 nm rms. For the monolithic 8-m mirror ATLAST concept, solid meniscus monolithic glass, as demonstrated on ground-based telescopes, requires no new technology, but will require engineering to ensure survival of launch vibrations/acoustics and the proper gravity sag unloading. For the segmented mirror ATLAST designs, hexagonal mirrors measuring 1.3 m and 2.4 m are baselined for the 9.2 m and 16 m apertures, respectively. Some potential options for the mirror composition include the Advanced Mirror System Demonstrator (AMSD) ULE glass segments developed for JWST, Actuated Hybrid Mirrors (AHM), and corrugated glass technology. ATLAST-9.2m has baselined the AMSD-like mirror segments of 1.3m size and 25 kg/m$^2$. At this mass and size, mirror segments are sufficiently stiff such that demonstration of better than 10 nm rms wavefront error and vibration and acoustic testing will be straightforward. An ATLAST technology development program would downselect between AMSD-like and AHM mirror segment technologies. Assuming both technologies meet launch loads and figure error, the downslect will be based on studies of thermal stability, cost, and manufacturing schedule. The TRL 6 milestone is a mirror segment that meets environmental and wavefront error budget requirements.

Wavefront sensing and control (WFS&C) also needs to be advanced from JWST to meet visible image quality with an allocation of 10-15 nm rms wavefront error for WFS&C residual. The segment's positions must be measured to a few nanometers at a rate faster than their support structure's time constants. Periodic image-based WFS employs phase retrieval to measure the wavefront error. For ATLAST-8m, this is done using WFS modules on each side of the three foci, updated every few weeks using stars of opportunity. For the segmented versions, this is done using several WFS modules that provide updates every few minutes (without interrupting science observations) using the guide stars simultaneously for wavefront sensing and guiding. Technology development requires a demonstration of WFS algorithms to the required performance within the processing capabilities of a flight computer with realistic guide star scenes. Actuators supporting the segments require a modest technology development to reduce the resolution to 2 nm and increase bandwidth. Since WFS&C is inherently a system-wide performance of the observatory, a demonstration of the telescope's operation would be planned with a subscale (~6 m class), partially populated (three-segment) testbed meeting error budgets under full thermal, vacuum, vibration, and jitter environments.

### 3.2 Detector Technology

Gigapixel detector arrays for visible imaging and ~500 Megapixel arrays for NIR imaging are required for studies of resolved stellar populations, galaxy evolution, and structure formation. Such arrays can be built with existing technology. However, development would result in better science performance (lower noise), lower risk (less complex electronics), lower cost, and lower power consumption. The wide field cameras in all ATLAST designs are envisioned to have ~1 gigapixel per channel. Exposure times invested in exoplanet and other faint object spectroscopy could be reduced by up to five times using photon counting detectors. Technology development of photon-counting CCDs is based on the low-light level CCDs built by E2V and the similar technology of Texas Instruments. The current TRL is 4; improvements in anti-reflection coatings, voltage swing, and charge induced clock noise would be demonstrated before the completion of a TRL 6 qualification program. In the longer term, CMOS based detectors are likely to be used.

UV detectors have ample room for improvements in efficiency and format size. UV spectroscopy, in particular, requires coatings, optics and detectors that are highly efficient. Current flight UV imaging detectors use CsI and $Cs_2Te$ photocathodes with 10% - 30% quantum efficiencies (QE). Photocathodes with QE of 30% - 80% using cesiated p-doped GaN have been produced in the lab, but have not yet been integrated into large format detector systems suitable for flight. Materials such as p-doped AlGaN and MgZnO should be developed for higher QE over a wider band, and better AR-coatings may be matched with p-channel radiation-hardened photon-counting CCDs. For the far UV, higher QEs also are required, and III/V semiconductor materials are effective. All must be coupled with large-format intensifier and readout systems. EBCCDs and microchannel plate methods (ceramic and glass) compete for the highest QE and largest formats, and both should be pursued.

### 3.3 Starlight Suppression Technology

An internal coronagraph or external occulter that provides $10^{-10}$ starlight suppression (output-to-input beam flux ratio) is required to characterize Earth-sized exoplanets. The best starlight suppression technology is not yet obvious. Fortunately, there are multiple options. The technology development plan addresses the viability of 1) a visible nulling coronagraph (VNC) that any telescope could use (development costs might be shared with large ground-based telescopes); 2) a Lyot-type coronagraph (for use with monolithic telescopes); and 3) an external starshade (a separate spacecraft creating a star shadow at the telescope). High contrast Lyot coronagraphs[39] are at TRL ~6 whereas the VNC and starshade are currently at or below TRL 4.

There are a number of coronagraphic configurations, each having tradeoffs between achievable contrast, IWA, throughput, aberration sensitivity, and ease-of-fabrication. Most techniques do not work on conventional (4-arm spider) obscured or segmented systems, but some non-conventional spider configurations (linear support) appear to allow performance competitive with off-axis systems. The VNC approach is a viable solution for *internal* suppression with a segmented telescope. VNC development is required to demonstrate $10^{-10}$ suppression ratios over at least a 23% passband. This requires the development of a spatial filter array (1027 fiber bundle), deformable mirror (MEMS 1027 segment), and an achromatic phase shifter.

Starshade theoretical performance has been validated by at least 4 independent algorithms and, in the lab, by two beamline testbeds[40]. Detailed CAD models exist for the New Worlds Observer[41] 50-meter starshade, which use high TRL components (membranes, hinges, latches, booms). For ATLAST, a larger starshade is required: 60 m to 80 m. The key challenges are primarily deployment reliability and shape control. The ATLAST starshade separation of ~165,000 km, while large, does not present any challenging formation flying or orbital dynamics issues but put additional requirements on the starshade propulsion system. The required starlight suppression performance with a segmented telescope is achievable with an external starshade[42]. The starshade technology developments are addressed through increasingly larger subscale models with TRL 6 being demonstrated through beamline tests, a half-scale quarter-section deployment, and a full-scale single petal deployment, performance, and environmental testing.

### 3.4 Rocket and Propulsion Technologies

All ATLAST concepts we investigated rely on the availability of a heavy lift launch vehicle with a high-volume fairing. The least massive of our ATLAST concepts (segmented 9.2m design) requires a payload-to-SEL2 capacity of ~16 metric tons. We assumed a launch of ATLAST-9.2m from KSC on an enhanced version of the Delta IV-Heavy launch vehicle. The enhancements, which are outlined in the Delta IV user's guide, consist of fairing with a 6.5 m outer diameter, a more powerful main engine (the RS68A which is already in testing) and the addition of 6 solid boosters already in use on the smaller version of the Delta IV. Informal estimates indicate that this enhanced version could carry 18 metric tons of payload mass to a C3 of about -0.69 $km^2/sec^2$, which is required for reaching SEL2. The most massive ATLAST concept (monolithic 8m) requires a payload-to-SEL2 capacity of ~55 metric tons and an inner fairing diameter of 8.8 meters. These requirements are consistent with the capabilities of the proposed Ares V launch vehicle (configuration LV 51.00.48). Cylindrical fairing segments of ~10 meters in height with an additional 10 – 12 meters of clearance over the central payload region in a conical or ogive segment are also required for our current ATLAST designs.

NASA's Evolutionary Xenon Thruster (NEXT) ion engine will achieve TRL 6 or greater through NASA's in-space propulsion programs. Such engines will be essential for efficiently enabling alignment of an external starshade with its companion observatory.

### 4. Concluding Remarks

The most compelling science cases for a future large-aperture UVOIR space telescope require major increases in angular resolution, sensitivity, and wavefront error stability over existing or planned facilities in the 0.11 μm to 2.4 μm wavelength range. ATLAST will definitively establish the frequency of detectable life on terrestrial-like exoplanets. ATLAST will revolutionize our understanding of star formation, galaxy evolution, and the nature of dark matter by its ability to provide unique and essential data to complement that from a new generation of very large ground-based optical and radio observatories anticipated to come on line in the 2010 – 2020 timeframe. Incremental advances in telescope aperture are not compelling, so radical advance is needed. Fortunately, the technology needs of several communities are aligning to make large space-based optical systems affordable. An affordable ATLAST observatory concept can satisfy the challenging science-driven performance requirements discussed here in time for a 2025-2030 launch if progress is made in maturing the key technologies needed.

Postman et al.: ATLAST Science Drivers[3] Andersen, D., et al., "*Performance Modeling of a Wide-Field Ground-Layer Adaptive Optics System*," PASP, 118, 1574-1590 (2006).
[4] Ammons, S.M., et al., "*First results from the UCSC Laboratory for Adaptive Optics multi-conjugate and multi-object adaptive optics testbed*," Proc. SPIE, 6272, (2006).
[5] Nicolle, M., Fusco, T., and Michau, V., Optics Letters, Vol. 29, Issue 23, pp. 2743-2745 (2004).
[6] Serabyn, G., Mawet, D., and Burruss, R., American Astronomical Society Meeting Abstracts, 216, #311.01 (2010).
[7] Guyon, O., et al., PASP, 122, 71, (2010).
[8] Postman, M., et al., *"Advanced Technology Large-Aperture Space Telescope (ATLAST): A Technology Roadmap for the Next Decade,"* arXiv:0904.0941, (2009). Report also on line at http://www.stsci.edu/institute/atlast
[9] Postman et al., *"Science with an 8-meter to 16-meter optical/UV space telescope,"* Proc. SPIE, 7010, (2008).
[10] Oegerle et al., *"ATLAST-9.2: A Deployable Large Aperture UVOIR Space Telescope,"* Bulletin of the American Astronomical Society, 41, 569 (2010).
[11] Stahl, H. P. et al., *"Design for an 8-meter monolithic UV/OIR space telescope,"* Proc. SPIE, 7436, (2009).
[12] Stahl, H. P., "*Survey of Cost Models for Space Telescopes*", Optical Engineering, Vol.49, No.05, (2010).
[13] Stahl, H. P. and Hinrichs, T., "*Multi-Variable Cost Model for Space Telescopes*", Proc. SPIE, 7731, (2010).
[14] Hickey, G. S., Lih, S.-S., Barbee, T. W., Jr., 2002, Proc. SPIE, 4849, 63, (2002).
[15] Strafford, D. N., DeSmitt, S. M., Kupinski, P. T., and Sebring, T. A., Proc. SPIE, 6273, 0R, (2006).
[16] Werner, M. et al., "*Active Space Telescope Systems – A New Paradigm,*" Astro2010 White Paper, (2010).
[17] Kasting, J., Traub, W. et al., arXiv:0911.2936, (2009).
[18] Lunine, J. I., et al., arXiv:0808.2754, (2008).
[19] Levine, M., et al., The Astronomy and Astrophysics Decadal Survey, 2010, 37, (2009).
[20] Guyon, O., Ap.J., 629, 592, (2005).
[21] Mountain et al., arXiv:0909.4503 (2009).
[22] Des Marais, D. J., et al., Astrobiology, 2, 153, (2002).
[23] Woolf, N. J., Smith, P. S., Traub, W. A., and Jucks, K. W., Ap.J., 574, 430, (2002).
[24] Ford, E. B., Seager, S., and Turner, E. L., Scientific Frontiers in Research on Extrasolar Planets, 294, 639, (2003).
[25] Brown, T. et al., "*The History of Star Formation in Galaxies*," Astro2010 Science White Paper, (2009).
[26] Oppenheimer, B. D., Davé, R., and Finlator, K., MNRAS, 396, 729, (2009).
[27] Birnboim, Y., and Dekel, A., MNRAS 345, 349, (2003).
[28] Keres, D., et al., MNRAS, 363, 2, (2003).
[29] Sembach, K. et al., "*The Cosmic Web*," Astro2010 Science White Paper, (2010).
[30] Bertone, S., Schaye, J., and Dolag, K., Space Science Reviews, 134, 295, (2008).
[31] Giavalisco, M. et al., "*The Quest for a Physical Understanding of Galaxies Across Cosmic Time*," Astro2010 Science White Paper, (2009).
[32] Sembach, K. et al., *"Technology Investments to Meet the Needs of Astronomy at Ultraviolet Wavelengths in the 21st Century,"* Astro2010 Technology Development White Paper, (2009).
[33] Savage, B. D., et al., Bulletin of the American Astronomical Society, 41, 502, (2010).
[34] Martin, N. F., Ibata, R. A., Chapman, S. C., Irwin, M., and Lewis, G. F., MNRAS, 380, 281, (2007).
[35] Simon, J. D., and Geha, M., ApJ, 670, 313, (2007).
[36] Strigari, L. E. et al., Nature, 454, 1096, (2008).
[37] Strigari, L. E., Bullock, J. S., and Kaplinghat, M., Ap.J., 657, L1, (2007).
[38] Soto, M., Kuijken, K., and Rich, R. M., IAU Symposium, 245, 347, (2010).
[39] Trauger, J. and Traub, W., Nature, 446, 771, (2007).
[40] Leviton, D. B., et al., Proc. SPIE, 6687, (2007).
[41] Cash, W., et al., Proc. SPIE, 7436, (2009).
[42] Soummer, R. et al., "*A Starshade for JWST: Science Goals and Optimization*", Proc. SPIE, 7440, 8, (2009).
12